\documentclass[twocolumn,showpacs,prl]{revtex4}
\usepackage{graphicx}
\usepackage{bm}
\begin{document}

\title{Formation of molecules in an expanding Bose-Einstein condensate}

\author{V. A. Yurovsky}

\author{A. Ben-Reuven}

\affiliation{School of Chemistry, Tel Aviv University, 69978 Tel Aviv,
Israel}

\date{\today}
\begin{abstract}
A mean field theory of expanding hybrid atom-molecule Bose-Einstein 
condensates is applied to the recent MPI experiments on ${}^{87}$Rb 
that demonstrated the formation of 
ultracold molecules due to Feshbach resonance. The subsequent dissociation 
of the molecules is treated using a non-mean-field parametric approximation.
The latter method is also used in determining optimal conditions for the 
formation of molecular BEC. 
\end{abstract}
\maketitle

Molecular Bose-Einstein condensates (BEC) have been recently formed in 
experiments on atomic BEC
\cite{HKMWCNG03,DVMR03,XMACMK03,MAXCK04} and on 
quantum-degenerate Fermi gases \cite{RTBJ03}. These experiments exploited 
the effect of Feshbach resonance \cite{TTHK99}, present when the energy
of an atomic pair is close to the energy of a bound molecular state.
The energy mismatch can be controlled by applying an external magnetic
field, thanks to the difference between the magnetic momenta of the 
molecule and of the atomic pair. 
By varying the magnetic field the energy of the molecular state is 
forced to cross
different states of the atomic pairs, belonging to a discrete spectrum 
in the case of a trapped gas, or to a continuum otherwise. The resonance 
value of the magnetic field strength $B_{0}$ corresponds to the crossing 
of the lowest discrete state, or the 
lower boundary of the continuum, respectively. 

In the experiments 
\cite{RTBJ03,HKMWCNG03,DVMR03,XMACMK03,MAXCK04} the resonances were
mostly swept in a backward direction, so that the molecular state
crossed the atomic ones downwards. This led to the transfer of population 
from the lowest 
atomic state in the case of BEC, or from an energy band in the case of 
a Fermi gas, to the molecular state, as had been proposed in Ref.\
\cite{MTJ00}. Assuming all the atomic population 
is initially in the BEC state, the backward sweep would have been ideally 
suitable for forming molecules, were it not for two destabilizing 
mechanisms. The resonant molecule is generally populated in an excited
rovibrational and electronic state, and therefore can be deactivated 
by inelastic collisions with atoms and other molecules (see Refs.\
\cite{TTHK99,YBJW99,YBJW00}). In addition, during the backward sweep 
some higher-lying non-condensate atomic states can be populated 
temporarily (by counterintuitive transitions --- see Ref.\ 
\cite{counter_int}), most notably in a strong resonance or at low 
densities (see Ref.\ \cite{YB03}). 
These two effects restrict the efficiency of conversion from the 
atomic condensate to the molecular one. 
The non-condensate atoms produced by molecular dissociation 
are formed as entangled atomic pairs (see Ref.\ \cite{YB03}). 
This process cannot be described by mean field (MF) theories. Therefore 
the analysis of formation of a molecular BEC incorporating both effects 
requires a non-MF approach in which the damping due to deactivating 
collisions is incorporated. Such an 
approach --- a generalized parametric approximation (PA) --- has been
described in Ref.\ \cite{YB03}, and applied to a Na BEC.
The non-MF effects become even more important in forwaed sweeps. 
An alternative study of molecular formation described in Ref.\
\cite{GKGTJ04}, using the macroscopic quantum-dynamics approach of Ref.\ 
\cite{KGB03}, takes into account non-MF effects but not inelastic 
collisions.

In the $^{87}$Rb experiments \cite{DVMR03} a BEC of $10^{5}$ atoms has 
been kept initially in a harmonic trap with frequencies 
$2\pi \times \left( 50, 120, 170\right) $ Hz. The magnetic field ramp 
was started, following the switching off of the trap, 
after a pre-expansion interval $t_{p}\ge 2$ ms. This measure allows the 
reduction of 
condensate density, and thus improves conversion efficiency (see Ref.\
\cite{YB03}). The field has been ramped from 1008 G to 1005.2 G with
various speeds, passing the 1007.4 G resonance of strength $\Delta
 \approx 0.2$ G
(see Ref.\ \cite{VDEMR03}), and held for 3 ms. During this time a
gradient magnetic field has been applied, leading to a relative motion
of the atoms and the molecules due to the difference of magnetic
momenta between the molecule and an atomic pair, $\mu \approx 2.8 \mu
 _{B}$  (see Ref.\ \cite{M02}). 
Next, the magnetic field was ramped up to 1008 G (in a forward sweep), 
converting the molecules back to atoms. The reconverted atoms inherit
the velocity of the molecules, and thus form a second condensate cloud.

\begin{figure}

\includegraphics[width=3.375in]{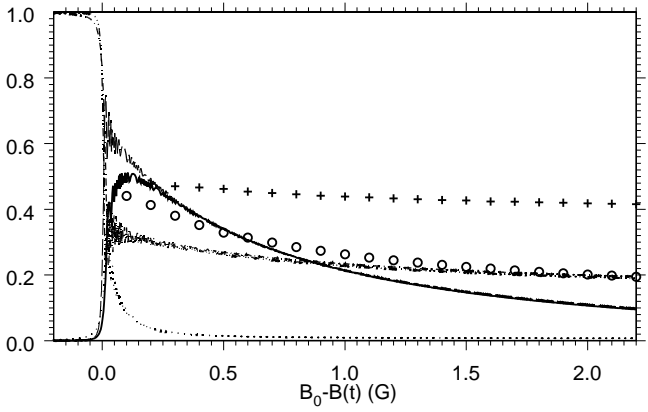}

\caption{Fraction of atoms surviving in the atomic condensate, and those 
converted to molecules and non-condensate atoms, calculated using the PA
(dot-dashed and solid lines, respectively) and the MF approach 
(dot-dot-dashed and dashed lines, respectively) for a homogeneous gas 
with the initial atomic density $3\times 10^{13}$cm$^{-3}$ and a 
magnetic ramp speed of 1 G/ms.
The fraction of non-condensate atoms calculated by using the PA is 
represented by the dotted line.
Pluses and circles represent the fraction of surviving and converted atoms,
respectively, calculated using the MF approach but taking into account 
the effects of spatial inhomogeneity and expansion.} 
\label{fig_mfpa}

\end{figure}

\paragraph{Association of atoms in an expanding BEC.} The PA approach 
as used in \cite{YB03} for a homogeneous gas  
is inapplicable directly to experiments on traps. 
However, the MF approach of \cite{YBJW99,YBJW00} (which can be easily 
extended to traps) should suffice for the description of association 
in the case concerned. This is shown by 
Fig.\ \ref{fig_mfpa}. It  compares the atomic and molecular densities
calculated with the PA approach \cite{YB03} and the MF approach of 
\cite{YBJW99,YBJW00}, for an initial atomic density corresponding to 
the peak density reached at the expansion time of 2.3 ms. 
The values used for the deactivation rate coefficients are 
$k_{a}=7\times 10^{-11}$  cm$^{3}/$s (see Ref.\ \cite{YB03b}) and
 $k_{m}=5\times 10^{-11}$  cm$^{3}/$s for
atom-molecule and molecule-molecule collisions, respectively, 
The elastic scattering length is $a_{a}=99$ atomic units (see Ref.\
 \cite{M02}).
As one can see, already when the magnetic field is 0.3 G
below the resonance (in a sweep totaling 2.2 G) the temporary 
occupation of non-condensate atom states practically vanishes, and the 
results of the two kinds of calculations coincide. 
For faster sweeps or higher densities the two results converge 
even faster.

The extension of the MF approach of Refs.\ \cite{YBJW99,YBJW00} 
to inhomogeneous expanding gases requires the solution
of two coupled Gross-Pitaevskii equations for the atomic 
$\varphi _{a}\left( {\bf r},t\right) $  and the 
molecular $\varphi _{m}\left( {\bf r},t\right) $ mean fields,
\begin{eqnarray}
i\hbar \dot{\varphi }_{a}\left( {\bf r},t\right) =\biggl\lbrack
 -{\hbar { } ^{2}\over 2m}\nabla ^{2}+\epsilon _{a}\left( t\right)
+V_{a}\left( {\bf r},t\right) -{i\over 2}k_{a}|\varphi _{m}\left(
 {\bf r},t\right) |^{2} \nonumber
\\
+{4\pi \hbar { } ^{2}\over m}a_{a}|\varphi _{a}\left( {\bf r},t\right
) |^{2}\biggr\rbrack \varphi _{a}\left( {\bf r},t\right) +2g^{
*}\varphi ^{*}_{a}\left( {\bf r},t\right) \varphi _{m}\left( {\bf
 r},t\right)  \nonumber
\\
{} \label{GP}
\\
i\hbar \dot{\varphi }_{m}\left( {\bf r},t\right) =\biggl\lbrack
 -{\hbar { } ^{2}\over 4m}\nabla ^{2}-{i\over 2}k_{a}|\varphi
 _{a}\left( {\bf r},t\right) |^{2} \nonumber
\\
-ik_{m}|\varphi _{m}\left( {\bf r},t\right)|^{2}\biggr\rbrack
 \varphi _{m}\left( {\bf r},t\right) +g\varphi ^{2}_{a}\left( {\bf
 r},t\right)  . \nonumber
\end{eqnarray}
Here $m$ is the mass of the atom and $\epsilon _{a}\left( t\right)
 =-{1\over 2}\mu \left( B\left( t\right) -B_{0}\right) $ is the
time-dependent Zeeman shift of the atom relative to half the energy of
the molecular state. The external magnetic field $B\left( t\right) $
 is kept constant
while $t<t_{0}$, and is linearly ramped at $t>t_{0}$ so that
 $B\left( 0\right) =B_{0}$. The atoms are considered 
trapped in a harmonic potential $V_{a}\left( {\bf r},t\right)
 ={m\over 2}\sum\limits^{3}_{j=1}\omega ^{2}_{j}r^{2}_{j}\theta \left
( t_{0}-t_{p}-t\right) $,
which is switched off before the ramping, at $t=t_{0}-t_{p}$. 
The atom-molecule coupling is
related to the phenomenological resonance strength $\Delta $ as
$|g|^{2}=2\pi \hbar ^{2}|a_{a}|\mu \Delta /m$ (see Ref.\
 \cite{YBJW00}).
The molecular trap potential and elastic collisions involving molecules
can be neglected since the molecules are formed after the expansion
starts when the trap is switched off and the densities decrease  
substantially.

The expansion of a pure atomic condensate cloud has been considered in
Ref.\ \cite{expan}. We generalize this theory here to the case of 
a hybrid atom-molecule condensate. 
Let us consider an initial atomic field with a Thomas-Fermi 
distribution, and a zero molecular field, at $t<t_{0}-t_{p}$. 
We can represent the two mean fields in the following form,
\begin{eqnarray}
\varphi _{a}\left( {\bf r},t\right) =A\left( t\right) \Phi _{a}\left(
 \bm{\rho },t\right) e^{i}{ } ^{S} \nonumber
\\
 \label{phi}
\\
 \varphi _{m}\left( {\bf
 r},t\right) =A\left( t\right) \Phi _{m}\left( \bm{\rho },t\right)
 e^{2i}{ } ^{S} ,\nonumber
\end{eqnarray}
using the scaled coordinates $\rho _{j}=r_{j}/b_{j}\left( t\right) $,
 $1\le j\le 3$, and a scaling factor
$A\left( t\right) =\left( b_{1}\left( t\right) b_{2}\left( t\right)
 b_{3}\left( t\right) \right) ^{-1/2}$. The scales $b_{j}$  obey
 the set of equations
\begin{equation}
\ddot{b}_{j}\left( t\right) =\omega ^{2}_{j}A^{2}\left( t\right)
/b_{j}\left( t\right) , b_{j}\left( t_{0}-t_{p}\right) =1 .
\end{equation}
The phase in Eq.\ (\ref{phi}),
\begin{equation}
S\left( t\right)  ={m\over \hbar}\sum\limits^{3}_{j=1}r^{2}_{j}{\dot{b}_{j}\left(
 t\right) \over 2b_{j}\left( t\right) }-{\epsilon_{0}\over\hbar}
 \int\limits^{t}_{t_{0}-t_{p}}dt^\prime A^{2}\left( t^\prime\right),
\end{equation}
accumulates most of the contributions of the kinetic energy. Here
$\epsilon _{0}=4\pi \hbar ^{2}a_{a}n_{0}/m$ is a chemical potential 
of the atomic BEC and $n_{0}$  is its
peak density while the trap is on.

Substitution of Eq.\ (\ref{phi})  into (\ref{GP}) leads to
the following set of coupled equations for the transformed mean fields,
\begin{eqnarray}
i\dot{\Phi }_{a}\left( \bm{\rho },t\right) &&=\left\lbrack \epsilon
 _{a}\left( t\right) -{i\over 2}A^{2}\left( t\right) k_{a}|\Phi
 _{m}\left( \bm{\rho },t\right) |^{2}\right\rbrack \Phi _{a}\left(
 \bm{\rho },t\right)  \nonumber
\\
&&+2A\left( t\right) g^{*}\Phi ^{*}_{a}\left( \bm{\rho },t\right)
 \Phi _{m}\left( \bm{\rho },t\right)  \nonumber
\\
{} \label{GPexp}
\\
i\dot{\Phi }_{m}\left( \bm{\rho },t\right) &&=-iA^{2}\left( t\right)
 \biggl\lbrack {1\over 2}k_{a}|\Phi _{a}\left( \bm{\rho },t\right)
|^{2} \nonumber
\\
&&+k_{m}|\Phi _{m}\left( \bm{\rho },t\right) |^{2}\biggr\rbrack \Phi
 _{m}\left( \bm{\rho },t\right) +A\left( t\right) g\Phi ^{2}_{a}\left
( \bm{\rho },t\right).  \nonumber
\end{eqnarray}
The residual kinetic energy terms are neglected here for the same
reasons as in Ref.\ \cite{expan}. The loss processes and atom-molecule
transitions distort $\Phi _{a}\left( \bm{\rho },t\right) $ and $\Phi
 _{m}\left( \bm{\rho },t\right) $ from the Thomas-Fermi
shapes, leading to additional energy shifts compared to the pure
atomic case. These shifts, however, are of the order of $A^{2}\left(
 t\right) \epsilon _{0}$, and can only 
lead to a negligible small shift of the resonance (of less than
 $10^{-4}$G in the present case).

Equations (\ref{GPexp}) have a rather clear physical sense. They
describe a ballistic expansion of the atomic and molecular BEC with
the same velocity distribution, reducing the densities by the 
factor $A^{2}\left( t\right) $, and
leading to a rescaling of the coupling and deactivation parameters. This
reflects the fact that the acceleration predates the formation of
molecules that inherit the velocity of the atoms they are formed
from.

\begin{figure}

\includegraphics[width=3.375in]{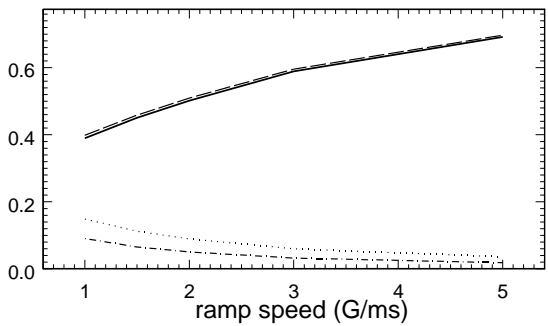}

\caption{Fraction of atoms surviving in the atomic condensate (solid
line) and those converted to molecules (dot-dashed line) calculated 
with the MF
approach, taking into account spatial inhomogeneity and expansion,
are shown for $k_{a}=10^{-10}$  cm$^{3}/$s. 
The dashed and dotted lines show the fraction of
surviving and converted atoms, respectively, calculated for
$k_{a}=7\times 10^{-11}$  cm$^{3}/$s.} \label{fig_ramp}

\end{figure}

The results of a numerical solution of Eq.\ (\ref{GPexp}) are
shown in Fig.\ \ref{fig_mfpa}. The inhomogeneity and the expansion
reduce the atomic density, leading to a slower loss of atoms and molecules.
The dependence on the sweep rate is presented in Fig.\ \ref{fig_ramp} for
two values of $k_{a}$  (the best fit and upper limit taken from Ref.\
\cite{YB03b}). The results are in agreement with the experimental data
of Ref.\ \cite{DVMR03} reporting that $\sim 7\%$ of atoms are converted to
molecules and $\sim 30\%$ remain in the atomic BEC for ramp speeds 
less than 2 G/ms.

\begin{figure}

\includegraphics[width=3.375in]{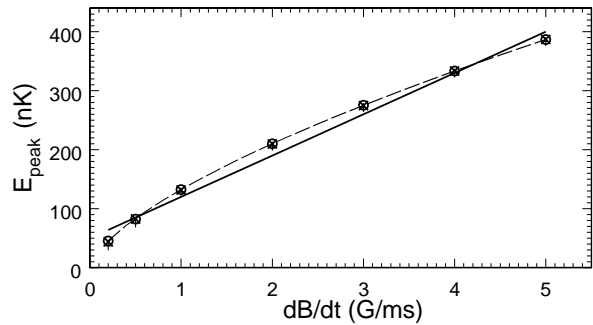}

\caption{Peak energy of the non-condensate atoms formed by the 
dissociation of molecules in a forward sweep calculated using Eq.\
(\protect\ref{Epeak}) (dashed line) and the PA, for the initial molecular
densities $10^{10}$  ($\circ$), $10^{11}$  ($\times $) and $10^{12}$
 (+)  cm$^{\text{-3}}$. The solid
line represents the results of a linear fit.} \label{fig_Epeak}

\end{figure}

\paragraph{Dissociation of molecules on a forward sweep.}
Consider now the dissociation of molecules on a forward sweep, when the
molecular state crosses the atomic ones upwards. Such a sweep has been used
in experiments for detection of molecules. The experiments demonstrate
no significant dependence of the reconversion efficiency on the ramp
speed. Our calculations using the PA \cite{YB03} demonstrate that under
the experimental conditions all molecules are dissociated except for a 
small part lost due to deactivating collisions during the sweep. The
molecules dissociate to entangled atomic pairs in a wide energy
spectrum (see Ref.\ \cite{YB03}) characterized
by the peak energy $E_{peak}$. The calculations demonstrate that 
$E_{peak}$ increases with the ramp speed (see Fig.\ \ref{fig_Epeak}). 
The results show no significant dependence on the molecular
density, justifying the applicability of the homogeneous density
approximation to real inhomogeneous situations. Although the ramp-speed 
dependence of $E_{\text{peak}}$ is clearly nonlinear, within the interval 
0.2 G/ms $<\dot{B}< 5$ G/ms it can be approximated by a straight line with 
the slope of 70 nK ms/G, which is close to the experimental value of 60 nK
ms/G reported in Ref.\ \cite{D03}.

In the present case 
the dissociation can be described by a simple analytical curve-crossing 
model based on
the approach of Ref.\ \cite{YBJW00}. A molecule dissociates into 
two entangled atoms, each with an energy $E$, by a crossing occurring at
$\mu \left( B\left( t\right) -B_{0}\right) =2E$. In the 
weak-resonance case, the quantum curve-crossing theory \cite{YBJ02} 
and the semiclassical Landau-Zener theory yield 
approximately the same result for the crossing probability, i.e., 
$e^{2\pi \lambda }-1\approx 1-
e^{-2\pi \lambda }\approx 2\pi \lambda $, where $\lambda =8\pi \hbar
 a_{a}n_{m}\left( t\right) \Delta /\left( m|\dot{B}|\right) $ (see
 Ref.\ \cite{YB03}).
Neglecting deactivating collisions the molecular density
$n_{m}\left( t\right) =|\varphi _{m}\left( t\right) |^{2}$ is given
 by the solution of Eq.\ (50) in Ref.\
\cite{YBJW00},
\[
n_{m}\left( t\right) =n_{m}\left( t_{d}\right) \exp\left\lbrack
 -2{a_{a}\mu \Delta \over \hbar { } ^{2}}\int\limits^{t}_{t
 _{d}}dt'\sqrt{m\mu \left( B\left( t'\right) -B_{0}\right)
 }\right\rbrack ,
\]
where $t_{d}$  is the starting time of the dissociation ramp. The
resulting energy distribution of the produced atoms is given by 
\begin{equation}
f\left( E\right) ={1\over 2}\sqrt{E}E^{-3
/2}_{\text{peak}}\exp\left\lbrack -{1\over 3}\left( E
/E_{\text{peak}}\right) ^{3/2}\right\rbrack ,
\end{equation}
with
\begin{equation}
E_{\text{peak}}={1\over 2m}\left( {\hbar ^{2}m|\dot{B}|\over
 4a_{a}\Delta }\right) ^{2/3} . \label{Epeak}
\end{equation}
This model is in good agreement with the
numerical results (see Fig.\ \ref{fig_Epeak}). Similar models have
been independently developed in Refs.\ \cite{MAXCK04,GKGTJ04}.

\begin{figure}

\includegraphics[width=3.375in]{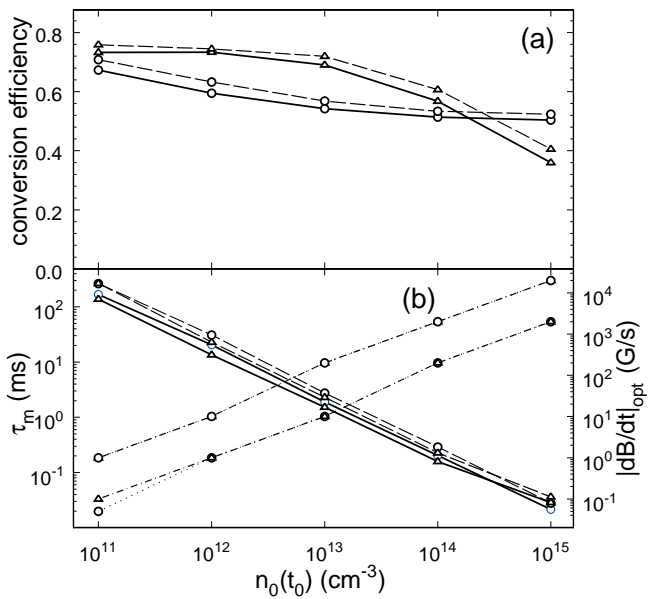}

\caption{Conversion efficiency (a) and the lifetime of the molecular
BEC $\tau _{m}$ (b) at the optimal ramp speed $\left( dB
/dt\right) _{\text{opt}}$, all  plotted as
functions of the initial atomic density $n_{0}$, calculated for 
the resonances at 1007 G 
(circles) and 685 G (triangles), using the rates of molecule-molecule
deactivation $10^{-10}$  (solid lines) and $0.5\times 10^{-10}$
 cm$^{3}/$s (dashed
lines). The dash-dotted and dotted lines in part (b) demonstrate
$\left( dB/dt\right) _{\text{opt}}$ for the two rates, respectively.} 
\label{fig_opt}

\end{figure}

\paragraph{Optimized molecular formation on a backward sweep.} Figure
\ref{fig_mfpa} demonstrates that the molecular population reaches a
maximum, corresponding to a conversion efficiency $\sim 50\%$, about
 0.1 G
below the resonance, and falls by half about 0.5 G below the
resonance. The low conversion efficiency observed in experiments
\cite{DVMR03} is due to collision loss during the long sweep to 2.2 G
below the resonance. Although the PA results of Fig.\ \ref{fig_mfpa}
overestimate the loss by neglecting the expansion, the molecular 
population of the expanding gas falls by half on reaching 2.2 G. Under the 
conditions of the experiments the lifetime of the molecules formed is 
about 0.5 ms, too short for an effective detection. 
As in the case of Na (see \cite{YB03}), 
the lifetime and conversion efficiency increase on reducing 
the initial condensate density. This is demonstrated in Fig.\ \ref{fig_opt},
showing results of calculations using the PA \cite{YB03} for the
homogeneous case. The use of this non-MF approach is important for such
situations, as the MF approach becomes inadequate in the vicinity of the peak 
molecular occupation (see Fig.\ \ref{fig_mfpa}), and at low condensate
densities. A use of the weaker resonance at 685 G with $\Delta \approx 17$
 mG $\mu =1.4 \mu _{B}$
(see Ref.\ \cite{M02}) should increase the conversion efficiency (see
Fig.\ \ref{fig_opt}).

\begin{figure}

\includegraphics[width=3.375in]{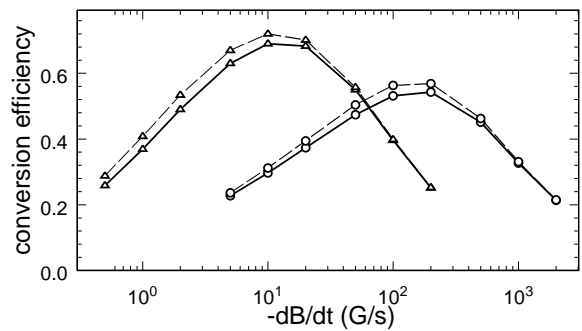}

\caption{Conversion efficiency as a function of the ramp speed $dB/dt$, 
calculated at the
initial atomic density $10^{13}$  cm$^{-3}$ for the resonances at 
1007 G (circles) 
and 685 G (triangles), using the rates of molecule-molecule deactivation
$10^{-10}$  (solid lines) and $0.5\times 10^{-10}$  cm$^{3}/$s
(dashed lines).}
\label{fig_conver}

\end{figure}

As in the case of Na studied in Ref.\ \cite{YB03}, the
conversion efficiency is determined by a concurrence of three
processes: the association of the atomic BEC and the two 
loss processes --- 
the dissociation of the molecular BEC onto non-condensate atoms and the 
deactivation by inelastic collisions. A reduction of the ramp speed 
enhances all three processes. 
The calculations of Ref.\ \cite{GKGTJ04}, taking no account
of the inelastic collisions, result in a monotonic increase of the
conversion efficiency with a decrease of the ramp speed. The inelastic
collisions, however, tend to reduce the conversion efficiency at slow
ramp speeds (see Fig.\ \ref{fig_conver}), as has also been demonstrated in
Ref.\ \cite{YB03}. The optimal ramp speed increases with the density
and the resonance strength (see Fig.\ \ref{fig_opt}).

\paragraph{Conclusions.} The experiments \cite{DVMR03} on formation of
molecules from a ${}^{87}$Rb BEC due to a Feshbach resonance
can be described by a MF theory of expanding atom-molecule BEC. Non-MF
calculations using the PA describe the energy of non-condensate atoms formed 
by molecular dissociation in a forward sweep and can predict 
optimal conditions for the formation of a molecular BEC in a 
backward sweep.

\end{document}